\documentclass{mn2e}
\usepackage{graphicx}
\usepackage{amsbsy}
\usepackage{amsmath}
\usepackage{epsf}

\newcommand{\beq}{\begin{equation}}
\newcommand{\eeq}{\end{equation}}
\newcommand{\msol}{{\rm M_{\odot}}}
\newcommand{\apjs}{ApJS}
\newcommand{\apj}{ApJ}

\newcommand{\mnras}{MNRAS}
\newcommand{\pasj}{PASJ}

\newcommand{\aap}{A\&A}
\usepackage{epstopdf}
\begin{document}
%
%===========================================================================
\title[Lognormal Column-Density Distributions]
{Do Lognormal Column-Density Distributions in Molecular Clouds Imply Supersonic Turbulence?}
%===========================================================================
%
\author[K. Tassis et al.]{K. Tassis$^{1}$, D. A. Christie$^{2}$, 
A. Urban$^{1}$, J. L. Pineda$^{1}$,  T. Ch. Mouschovias$^{2}$, 
H. W. Yorke$^{1}$, \newauthor H. Martel$^{3,4}$
\\$^1$Jet Propulsion Laboratory, California Institute of Technology, Pasadena, CA 91109, USA
\\$^2$Departments of Physics and Astronomy, University of Illinois at Urbana-Champaign, 
1002 W. Green Street, Urbana, IL 61801, USA
\\$^3$D\'{e}partement de Physique, de G\'{e}nie Physique et d'Optique, 
Universit\'{e} Laval, Qu\'{e}bec, QC, G1K7P4, Canada
\\$^4$Centre de Recherche en Astrophysique du Qu\'{e}bec }
\maketitle
\label{firstpage}
\begin{abstract}

Recent observations of column densities in molecular clouds find lognormal 
distributions with power-law high-density tails. These results are
often interpreted as indications that supersonic turbulence
dominates the dynamics of the observed clouds. 
We calculate and present the column-density distributions of three clouds, 
modeled with very different techniques, 
none of which is dominated by supersonic turbulence.  
The first star-forming cloud is simulated using smoothed particle hydrodynamics (SPH); 
in this case gravity, opposed only by thermal-pressure forces, drives the evolution.
The second cloud is magnetically subcritical with subsonic turbulence, simulated 
using nonideal MHD; in this case the evolution is due to gravitationally-driven 
ambipolar diffusion. The third cloud is isothermal, self-gravitating, and has  
a smooth density distribution analytically approximated
with a uniform inner region and an $r^{-2}$ profile at larger radii. 
We show that in all three cases the column-density distributions are lognormal.
Power-law tails develop only at late times (or, in the case of the smooth analytic profile, 
for strongly centrally concentrated configurations), when gravity dominates all opposing forces. 
It therefore follows that lognormal column-density distributions are generic features of 
diverse model clouds, and should not be interpreted as being a consequence of supersonic turbulence.
\end{abstract}


%---------------------
\section{Introduction}\label{sec:intro}
%---------------------

Recent observations indicate that the column-density distribution in molecular 
clouds is typically lognormal. For example,  Goodman et al.\ (2009) studied  
the Perseus molecular cloud using 2MASS, IRAS, and $^{13}$CO data, and found 
that the column-density distributions of both the cloud as a whole and that of smaller, 
few-parsec ``subregions'' are lognormal. Kainulainen et al.\ (2009) probed 
the column-density distributions in 23 molecular cloud complexes using near-infrared dust 
extinction maps from the 2MASS data archive, also finding lognormal column density 
distributions, with power-law tails present in star forming clouds. 
Lombardi, Alves, \& Lada (2006) analyzed 2MASS data for the Pipe nebula and found 
that the column-density distributions are complex; however because multiple velocity 
components are present in C$^{18}$O (Onishi et al.\ 1999; Muench et al.\ 2007), 
the complexity may be the effect of superposition of more than one cloud observed in 
projection, and the underlying individual components may still be consistent with lognormal 
distributions. Wong et al.\ (2008) observed the column-density distribution of the Giant 
Molecular Cloud RCW 106 in $^{13}$CO and found it to be lognormal. Finally, Pineda et al. 
(2010) also found lognormal column-density distributions with tails for the Taurus molecular 
cloud using 2MASS extinction and CO observations.

These results have been attributed to the dominance of supersonic turbulence 
in the observed regions. When gas turbulent pressure dwarfs the thermal-pressure term
with negligible gravitational and magnetic forces, 
the hydrodynamic equations become scale invariant, and the  distribution 
of the volume density accordingly becomes lognormal (V\'{a}zquez-Semadeni 1994). 
Ostriker et al. (2001) have shown that, in highly supersonic turbulent ideal-MHD simulations 
including self-gravity, the column-density distribution is also lognormal.
Vazquez-Semadeni \& Garc\'{i}a (2001) examined turbulent 3-dimensional simulations, 
both magnetic and nonmagnetic, as well as random realizations of 3-dimensional 
lognormal density fields. They argued that the shape of the column-density distribution 
depends on the number of correlation lengths included in the line-of-sight over which 
the density integration occurs. As the integration path increases from less than one to 
many correlation lengths, the column-density distribution transitions from lognormal 
(the underlying 3-dimensional density field distribution) to exponential, to Gaussian 
(due to the central limit theorem, when many uncorrelated patches are summed up along 
the line of sight). They proposed to invert the argument to derive from the observed 
shape of the column-density distribution the number of correlation lengths 
in clouds along the line of sight. They suggested that the reason for which column density 
distributions are found to be lognormal with exponential tails in simulations is that too few
correlation lengths are included to render the distributions Gaussian. 
Federrath et al.\ (2009) find, in purely hydrodynamic isothermal simulations 
with supersonic turbulence and no self-gravity, lognormal volume- and column-density 
distributions, the width of which depends on the driving mechanism -- since these are 
driven turbulence simulations. Small deviations in the wings, mostly in the low-density wing, 
were attributed to intermittency of the velocity field.

\begin{figure}
\begin{center}
\includegraphics[width=3.0in]{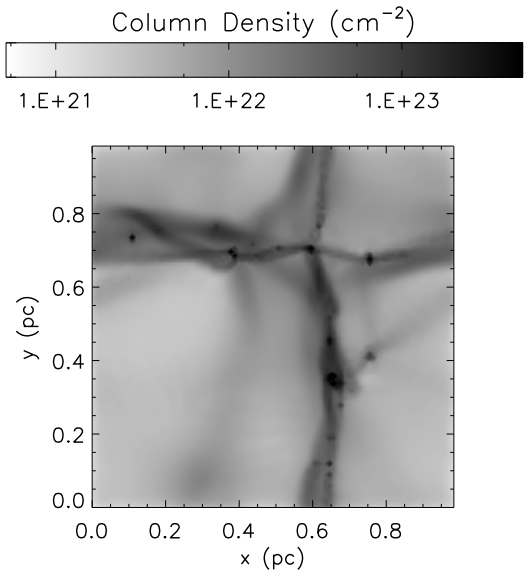}
\caption{Column-density map of simulation A at $t=0.72$ Myr. 
The grayscale corresponds to the (logarithmic) column-density scale. }
\label{AndreaPic}
\end{center}
\end{figure}
\begin{figure}
\begin{center}
\includegraphics[width=3.0in]{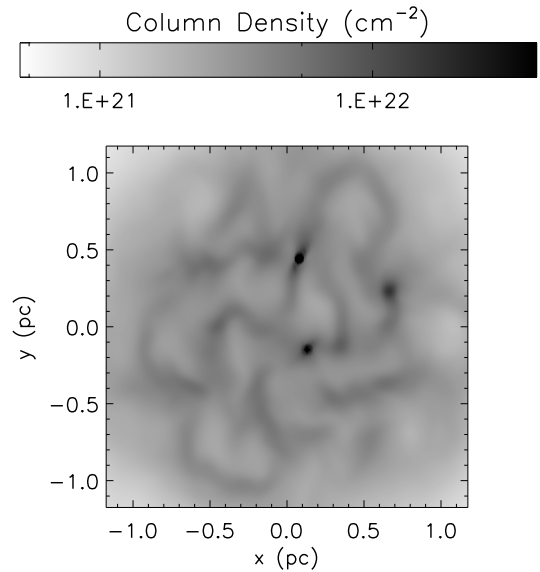}
\caption{Column-density map of simulation B at $t=4.8$ Myr. 
The grayscale corresponds to the (logarithmic) column-density scale. }
\label{DuncanPic}
\end{center}
\end{figure}

Although these studies establish that model molecular clouds dominated by supersonic 
turbulence exhibit lognormal distributions of column densities, they do 
not demonstrate that the dominance of supersonic turbulence is the prime cause of such 
column-density distributions or even that supersonic turbulence is a necessary ingredient.
In other words, there may be other features
of the assumed model clouds that are responsible for this effect. {\it Even if} 
dominance of supersonic turbulence is a {\it sufficient} condition to produce lognormal
column-density distributions, it does not follow that it is also a {\it necessary}
condition.

In this paper we test the uniqueness of the supersonic-turbulence interpretation and
study the ubiquity of lognormal column-density distributions. In particular, we 
investigate whether lognormal column-density distributions can also be produced in 
model molecular clouds {\em not} dominated by supersonic turbulence. We do so by 
examining the column-density distributions in three very different examples of model 
molecular clouds. In the first example, the molecular cloud is simulated using Smoothed 
Particle Hydrodynamics (SPH, Monaghan 1992), including self-gravity and radiative heating/cooling.
In the second example, a 3-dimensional, magnetically subcritical molecular cloud evolves 
due to gravitationally-driven ambipolar diffusion in the presence of {\em subsonic} turbulence. 
In the third example, we examine the analytically approximated density distribution of an isothermal, 
self-gravitating cloud having a uniform-density central region and a power-law profile 
at larger radii (Dapp \& Basu 2009). 

In \S\,\ref{dsims} we present the basic relevant properties of the three model clouds and 
discuss the evolution of the first two -- the third one is a static model. In \S\,\ref{theresults} 
we present and discuss the column-density distributions in each of the model clouds, and their 
dependence on time or on profile parameters as appropriate. The conclusions are summarized 
in \S\,\ref{conc}.

\section{The Model Clouds}
\label{dsims}

The first simulation (simulation A) is described in detail in Urban et al.\ (2010a, 2010b). 
Here we briefly summarize its main features. It uses the SPH algorithm with particle splitting
and sink particles (Martel et al.\ 2006). The simulation box is approximately 1 pc$^3$ in volume 
with periodic boundary conditions.  It contains a total mass of 670 M$_{\odot}$.  
The material is initially distributed 
uniformly with small density perturbations that reproduce a Gaussian random field with a density power 
spectrum $P(k) \propto k^{-2}$. The initial density of the region is $n=1.22 \times 10^4$ cm$^{-3}$ 
and the initial temperature is $T=5$ K. Sink particles (which can be interpreted as single stars or groups of stars) 
form at a density $n=7.3 \times 10^{7}$cm$^{-3}$ or mass 0.008 M$_{\odot}$. The simulation is stopped when 
the highest sink-particle mass reaches 21 M$_{\odot}$.  In addition to gravity, the simulation also 
includes heating and cooling of both gas and dust. The radiative effect of forming stars is included 
using the algorithm described in Urban et al.\ (2009). The luminosity from the young stars (sink particles) heats the dust, which is collisionally coupled to the gas. Molecular cooling and heating from cosmic-ray ionisation is also included in the calculation of the gas temperature. A column-density map of simulation A at $t=0.72$ Myr, which is the final time, is shown in Fig. \ref{AndreaPic}.

The second simulation (simulation B) is an MHD run using Zeus-MP (Hayes et al. 2006) which has been 
extended to include ambipolar diffusion. The simulation box is 8 pc $\times$ 8 pc $\times$ 2 pc with the third dimension being the direction of the initial 16.3 $\mu$G magnetic field. The MHD boundary 
conditions are reflective and the boundary conditions on the gravitational field are appropriate 
for an isolated cloud. The initial density distribution is $n=300$ cm$^{-3}$ within a cylindrical 
radius of 2 pc, and with a Gaussian tail beyond that\footnote{This low-density ``envelope'', beyond the inner 2 pc region of the model cloud, is added for computational convenience (to avoid reflection of waves).  It does not have any physical significance, so it is not shown in the display of the physical results. The results are not affected by its presence, provided that its size is large enough.}. The initial central mass-to-(magnetic)flux ratio is 0.9 times the critical value for collapse. The cloud is allowed to relax to equilibrium before a Gaussian random velocity field (with subsonic root-mean-square velocity dispersion $\simeq 0.5$ $C_{\rm s}$ (where $C_{\rm s}$ is the isothermal sound speed) is introduced and ambipolar diffusion is turned on. The simulation is stopped when the maximum density reaches $5 \times 10^6$ cm$^{-3}$. The column-density map at this time ($t=4.8$ Myr), looking down the initial direction of the magnetic field, is shown in Fig. \ref{DuncanPic}. At this time, there are eighteen identifiable self-gravitating cores, separated by a mean distance $\simeq$ 0.3 pc, and with a range of masses 0.32 - 15 $\msol$. More details for this run and other, similar runs, but with different input paramaters, are given in Christie \& Mouschovias (2010).

\begin{figure}
\begin{center}
\includegraphics[width=3.0in]{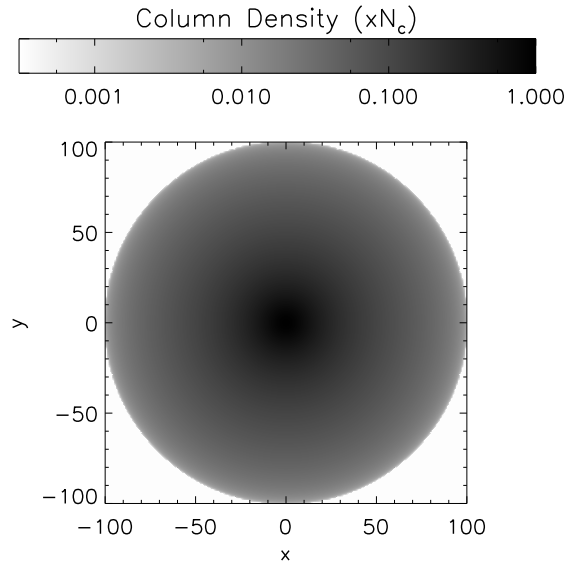}
\caption{Column density of a cloud with a smooth density profile given by Eq. (\ref{rhoeq}) and $(a,c)=(10,10)$.}
\label{SmoothPic}
\end{center}
\end{figure}

The third case we examine is an approximate, analytic, {\em smooth} density profile,
without any random perturbations. The analytic expression is designed to fit a spherical, 
isothermal cloud, with a flat inner region and a power-law density profile at larger radii 
(a Bonnor-Ebert sphere, with only thermal pressure opposing gravity, and not necessarily in equilibrium):
\begin{equation}
\label{rhoeq}
\rho(r) = \left\{\begin{array}{ll}\rho_ca^2/(r^2+a^2) & r\leq R \\0 & r>R\end{array}\right.\,,
\end{equation}
where $\rho_{\rm c}$ is the central value of the density, $R$ is the radius of the cloud, 
and the parameter $a$ determines the size of the uniform-density inner region and is 
proportional to the Jeans length. The column density for this cloud is then (Dapp \& Basu 2009)\
\begin{eqnarray}
N(\tilde{r}) &=& \frac{N_{\rm c}}{\sqrt{1+(\tilde{r}/a)^2}}\nonumber \\
 & & \times \left. \left. \left[\arctan \left(\sqrt{\frac{c^2-(\tilde{r}/a)^2}{1+(\tilde{r}/a)^2}}\right) 
 \right/ \arctan(c)\right]\,, \right.
\end{eqnarray}
where $N_{\rm c}$ is the central value of the column density, $\tilde{r}$ is the radial coordinate 
on the plane of the sky, and $c=R/a$, the size of the cloud relative to that of the uniform-density 
central region. We sample this profile with a 1024$\times$1024 grid, and we then construct a  
distribution of the column-density values within the grid.  We  examine two cases of 
this profile. 
The first has $(a,c)=(10,10)$ and corresponds to a cloud with a substantial flat inner 
region compared to the total volume of the cloud. The central gas density is 101 times 
greater than the density at the surface of the cloud. A column-density map of this cloud 
is shown in Fig. \ref{SmoothPic}. The second case has $(a,c)=(2,350)$ and represents a much more 
centrally concentrated cloud (in the limit $a\rightarrow 0, c\rightarrow \infty$ the profile becomes 
a singular isothermal sphere). For the second case the central gas density is $\sim 10^5$ times 
greater than the density at the surface of the cloud.

\begin{figure*}
\begin{center}
\includegraphics[width=6.0in]{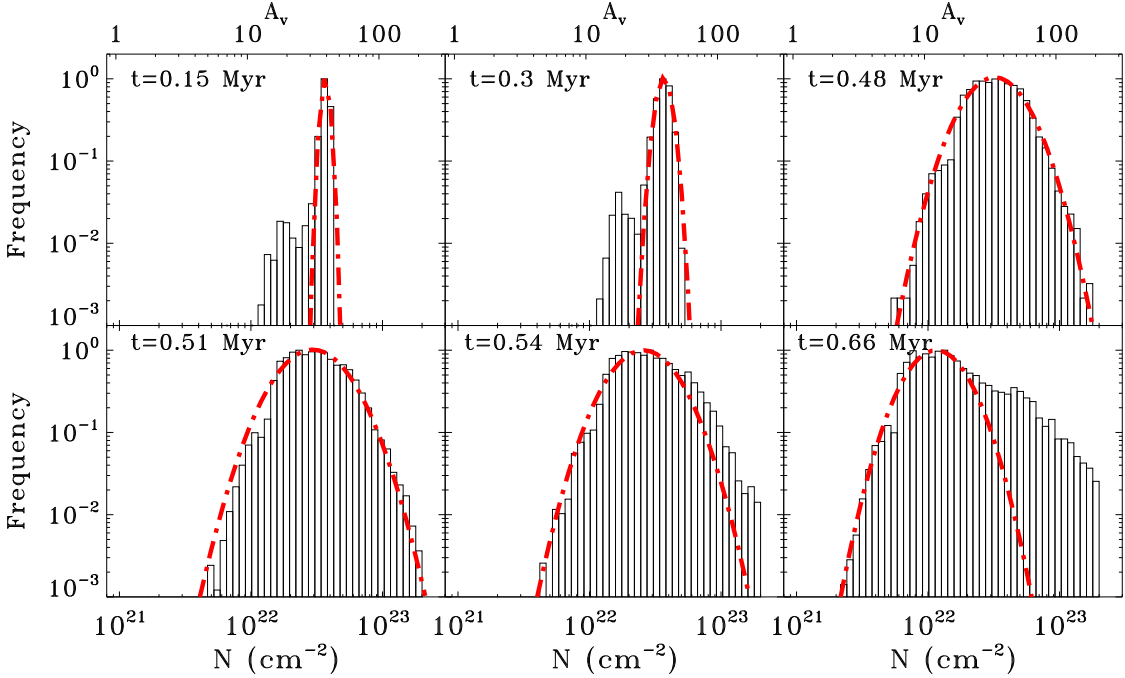}
\caption{Snapshots at different times of the column-density distribution in a molecular 
cloud, whose evolution is followed using an SPH code (simulation A). 
The evolution is initiated and controlled by self-gravity, and proceeds at the free-fall timescale. 
The dot-dashed line is the 
best fit lognormal that describes the histograms. The high column-density tail 
develops at late times due to the formation of cores and protostars inside the molecular cloud. }
\label{simA}
\end{center}
\end{figure*}

\begin{figure*}
\begin{center}
\includegraphics[width=6.0in]{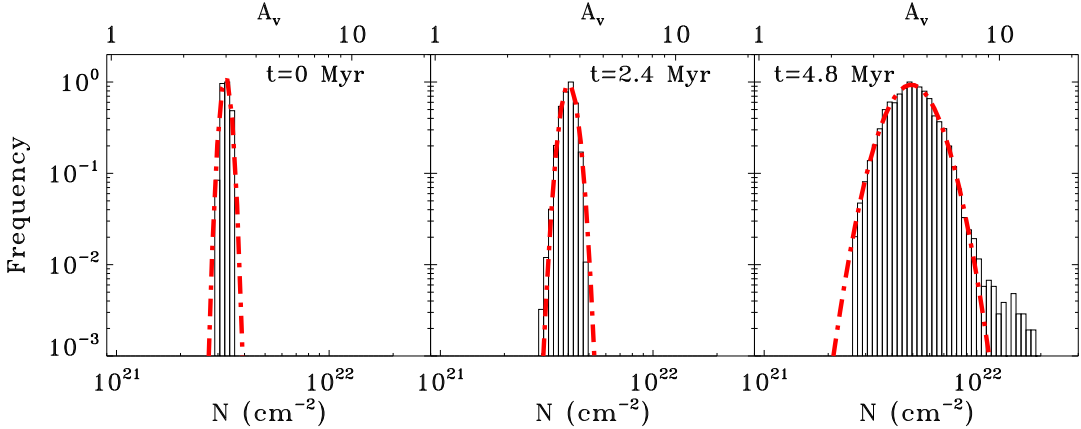}
\caption{Snapshots at different times of the column-density distribution in a molecular cloud, 
whose evolution is followed using a 3D, nonideal MHD code (simulation B). The evolution observed 
here is due to gravitationally-driven ambipolar diffusion. The dot-dashed line is the best fit 
lognormal that describes the histograms. }
\label{simB}
\end{center}
\end{figure*}

\begin{figure*}
\begin{center}
\includegraphics[width=6.in]{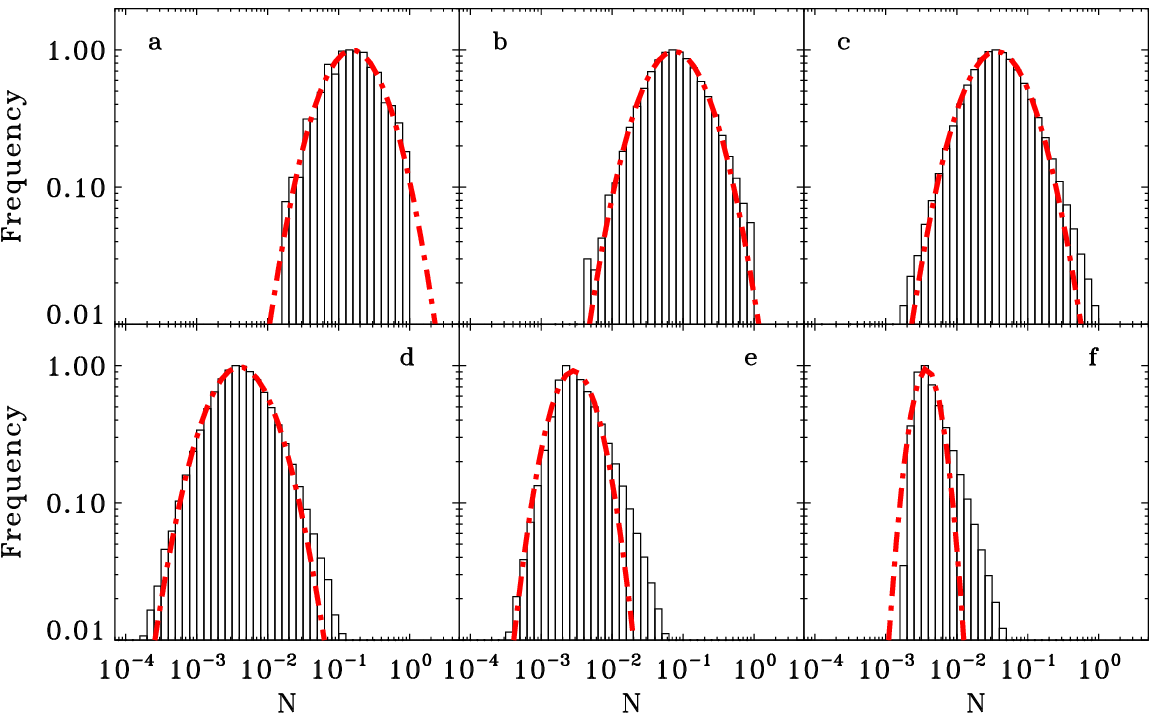}
\caption{Column-density distribution for a cloud with a smooth density profile given by 
Eq. (\ref{rhoeq}). {\it Upper row}: cloud with an extended uniform inner region and a power-law
``envelope'' [$(a,c)=(10,10)$]. {\it Lower row}: 
very centrally concentrated cloud, with a very small uniform inner region and a power-law 
``envelope'' [$(a,c)=(2,350)$]. The middle panels in each row show the column-density distribution 
when the clouds are sampled on a 1024$\times$1024 grid.  The left (right) panels show the column-density 
distribution of the same clouds, when these are sampled with grid resolution 
512$\times$512 (2048$\times$2048).
The upper-row cloud exhibits an almost perfect lognormal column-density distribution; in the lower-row cloud   the column-density distribution is also lognormal but, 
in adddition, has a power-law tail. }
\label{caseC}
\end{center}
\end{figure*}

\section{Results}
\label{theresults}

The results for simulations A and B are summarized in Figs. \ref{simA} and \ref{simB} in which we 
show snapshots at different times of the column-density distribution (represented in the form of a histogram) 
in each simulation. The time corresponding to each snapshot is indicated as a label in each panel. 
The lower horizontal axis shows the column density in units of ${\rm cm^{-2}}$ while the upper axis 
shows extinction (A$_{\rm V}$) using the conversion $N_{\rm H_2}/A_{\rm v} = 9.4\times 10^{20}{\rm cm^{-2}/ mag}$ (Bohlin et al.\ 1978). For simulation A, the column density is sampled with a $100\times 100$ grid, 
using 51 logarithmically-spaced  bins, spanning column density values from $1.5\times 10^{20}$ to $2\times10^{23}$  ${\rm cm}^{-2}$. For simulation B, the distribution of column densities in the central $2.3\times 2.3$ pc is sampled using again 51 logarithmically spaced  
bins, now spanning column density values from  $8\times 10^{20}$ to $2\times10^{22}$  ${\rm cm}^{-2}$. In this region, the simulation grid is uniform and it has $118\times 118$ cells. 
The velocity dispersions increase in each snapshot with increasing time, from 
0.48$C_{\rm s}$ at $t=0$ to 0.55$C_{\rm s}$ at $t=2.4$ Myr and to $0.58 C_{\rm s}$ at $t=4.8$ Myr, 
but remain subsonic throughout the evolution. The dot-dashed line is the best-fit lognormal distribution for each histogram. Both the horizontal and the vertical axes are logarithmic (the latter being the frequency of incidence for each column-density value normalized so that the most frequent value has a frequency of 1), so the lognormal distribution has a parabolic shape. In both cases, as time progresses and density peaks develop, the width of the column-density distribution increases, but not as a result of successive compressions and rarefactions due to supersonic turbulence. The shape of the distribution remains lognormal, and a power-law tail develops at late times and at high densities. In simulation A, the power-law tails in the lognormal distribution are more pronounced because of the use of sink particles that allow the simulation to continue to more advanced evolution stages, while simulation B is stopped soon after a few cores form. Power-law tails in simulation A start to appear when the fraction of mass in sink particles becomes non-negligible (3\% of the mass at $t=0.54$ Myr) and they become more pronounced as the fraction of the mass in sink particles increases (30\% at $t=0.66$ Myr, corresponding to the lower right panel of Fig. \ref{simA} with the most significant power-law tail). At low column densities, deviations from the lognormal shape are present in certain snapshots of simulation A. They appear at early times but correspond to a very small fraction of the gas in the simulated cloud. Similar deviations from the lognormal shape are also evident in observed clouds (Kainulainen et al. 2009).

The evolution is generally slower in the nonideal MHD run, since the cloud is initially magnetically subcritical, and density peaks  develop only after ambipolar diffusion has had enough time to create magnetically supercritical fragments within the cloud. The simulated cloud presented here is purposely selected to be subcritical to demonstrate that, even if the magnetic forces are the dominant opposition to gravity, the formation of lognormal column-density distributions is not inhibited. The qualitative evolution of the simulation does not change with varying the initial mass-to-flux ratio; only the evolutionary timescale changes. In the case of the SPH simulation A the evolution is much faster and proceeds on a free-fall timescale. Self-gravity dictates the evolution of the simulated cloud and is responsible for the shape of the distribution of column densities, not just its deviation from a lognormal shape at later times.

In case C, after the smooth column-density field is sampled on the 1024$\times$1024 grid, a column density distribution is constructed using  41 bins, logarithmically spaced from $10^{-5}$ to $10^{1}$ (in units of the central column density). The column-density distributions for  the cloud with a substantial uniform inner region [$(a,c)=(10,10)$] and for the almost singular cloud  [$(a,c)=(2,350)$] are shown in the upper and lower row middle panels (panel b and panel e) of Fig. \ref{caseC}, respectively. Even in this case, in which random perturbations of the density field are {\em completely absent}, a lognormal function fits remarkably well the distribution of the column densities. In the case of the more centrally concentrated cloud, we can also see the presence of a power-law high-density tail. In the limit of a singular isothermal-sphere profile, the column-density distribution asymptotes to a pure power-law. 

We have also tested the effect of the spatial resolution of the grid, with which a column-density map
is sampled, on the resulting distribution of column densities. For case C, we explicitly plot 
column-density distributions resulting from grids of different spatial resolutions. 
In each row of Fig. \ref{caseC}, the leftmost and rightmost panels show the same column-density distributions of the clouds as in the middle panel, but sampling occurs with grids of half and twice the spatial resolution, respectively. For the cloud with a substantial uniform inner region (upper row), the grid resolution  does not affect the shape of the distribution, which is, in all cases, well fitted by a lognormal. With increasing resolution, the distribution simply becomes less noisy and, as expected for an underlying density profile with a power-law tail, the most frequent column-density value decreases slightly. However, for the more centrally concentrated cloud (lower row), decreasing grid resolution leads to suppression of the power-law tails and a distribution shape better fitted by a lognormal. We have also tested the effect of smoothing down the outputs of simulations A and B to decreased spatial resolutions, and we have found that the resulting column-density distributions do not show any qualitative change - the shape of the distributions remains lognormal, with power-law tails at late times. 

\section{Discussion and Conclusions}
\label{conc}

The goal of this investigation has been to test whether the frequently adopted interpretation of the often observed lognormal distribution of column densities in molecular-cloud regions, i.e, that supersonic turbulence dominates the cloud dynamics, is unique or necessary. To this end, we have studied the distributions of the column density in three very different classes of model clouds: a compact and dense cloud simulated through SPH, with the evolution dominated by gravity; an isothermal, magnetically subcritical cloud with subsonic random initial velocity perturbations, with the evolution controlled by gravitationally-driven ambipolar diffusion; and a cloud with a smooth density profile, representing an isothermal, self-gravitating spherical object with a uniform inner region and a power-law profile at large radii, without any random perturbations. 

We have shown than  {\em in all cases} the column-density distributions are lognormal, with power-law tails 
developing at late times in simulations A and B, or, in the case of the smooth analytic profile, for very centrally concentrated configurations. {\em The lognormal shape of column-density distributions is not due to supersonic turbulence in any of the cases studied}, and the power-law tails are not due to intermittency or gravity taking over the dynamics. In both simulations studied here, the clouds have been self-gravitating and the evolution gravity-driven from the start, even before the development of tails. The tails develop as one or more strong density peaks appear in the simulated clouds. 

Furthermore, we have tested the effect of the resolution of the grid used to sample the column density of a cloud. We have found that for an intrinsically lognormal column-density distribution, the grid resolution does not have a significant effect on the observed distribution; however, when power-law tails are present, a poor sampling resolution can suppress the tails and result in a distribution that is better fitted by a lognormal shape. 

The results of this study highlight the fact that the distribution of column densities in gas clouds is a statistical characterization of the medium, and as such it does not have a one-to-one correspondence with the underlying physics that governs the dynamics of the cloud. Instead, we have found that lognormal column density distributions are a natural outcome of the evolution of a molecular cloud regardless of whether turbulence, gravity, or ambipolar diffusion in magnetically subcritical clouds initiates or determines the evolution.

\section*{Acknowledgments}

We are grateful to Neal Evans for enlightening discussions.  TM's work is partially supported by the National Science Foundation under grant NSF AST-07-09206 to the University of Illinois. HM is supported by the Canada Research Chair program and NSERC. This work made extensive use of the NASA Astrophysics Data System and {\tt arXiv.org} preprint server. AU's research was partialy supported by an appointment to the NASA Postdoctoral Program at the Jet Propulsion Laboratory, administered by Oak Ridge Associated Universities through a contract with NASA. Part of this work was carried out at the Jet Propulsion Laboratory, California Institute of Technology, under a contract with the National Aeronautics and Space Administration.  

\copyright 2010. All rights reserved.


\begin{thebibliography}{}
%
%CT: I alphabetized the list; the first two references were out of order.
%
\bibitem[Bohlin et al.(1978)]{1978ApJ...224..132B} Bohlin, R.~C., Savage,
B.~D., \& Drake, J.~F.\ 1978, \apj, 224, 132
\bibitem[Dapp \& Basu(2009)]{2009MNRAS.395.1092D} Dapp, W.~B., \& Basu, S.\ 2009, 
 \mnras, 395, 1092 
\bibitem[Federrath et al.(2009)]{federrath09} Federrath, C., 
Roman-Duval, J., Klessen, R., Schmidt, W., 
\& Mac Low, M.~-M.\ 2009, arXiv:0905.1060 
\bibitem[Goodman et al.(2009)]{goodman09} Goodman, A.~A., Pineda, 
J.~E., \& Schnee, S.~L.\ 2009, \apj, 692, 91 
\bibitem[Hayes et al.(2006)]{2006ApJS..165..188H} Hayes, J.~C., Norman, 
M.~L., Fiedler, R.~A., Bordner, J.~O., Li, P.~S., Clark, S.~E., ud-Doula, 
A., \& Mac Low, M.-M.\ 2006, \apjs, 165, 188 
\bibitem[Kainulainen et al.(2009)]{kainulainen09} Kainulainen, J., 
Beuther, H., Henning, T., \& Plume, R.\ 2009, arXiv:0911.5648 
\bibitem[Lombardi et 
al.(2006)]{Lombardi06} Lombardi, M., Alves, J., \& Lada, C.~J.\ 2006,
\aap, 454, 781 
%
%CT: Add Martel et al. 2006 here
%TM: Done
\bibitem[Martel et al.(2006)]{2006ApJS..163..122M} Martel, H., Evans, 
N.~J., II, \& Shapiro, P.~R.\ 2006, \apjs, 163, 122 
\bibitem[Monaghan(1992)]{M1992} Monaghan, J.~J.\ 1992, ARA\&A, 30, 543 
\bibitem[Muench et al.(2007)]{muench07} Muench, A.~A., Lada, 
C.~J., Rathborne, J.~M., Alves, J.~F., 
\& Lombardi, M.\ 2007, \apj, 671, 1820 
\bibitem[Onishi et al.(1999)]{1999PASJ...51..871O} Onishi, T., et al.\ 
1999, \pasj, 51, 871 
\bibitem[Ostriker et al.(2001)]{2001ApJ...546..980O} Ostriker, E.~C., 
Stone, J.~M., \& Gammie, C.~F.\ 2001, \apj, 546, 980 
\bibitem[Pineda et al. (2010)]{pineda2010} Pineda, J.L., 
Goldsmith, P., Chapman, N., Snell, R.L., Li, D., Cambr\'{e}sy, L., \& 
Brunt, C. 2010, {\em submitted to ApJ}
\bibitem[Urban et al.(2009)]{2009ApJ...698.1341U} Urban, A., Evans, N.~J., II, 
\& Doty, S.~D.\ 2009, \apj, 698, 1341
\bibitem[Urban et al.(2010a)]{U2010a} Urban, A., Martel, H., \& Evans,
N.~J., II, \ 2010a, \apj, 710, 1343
\bibitem[Urban et al.(2010b)]{U2010b} Urban, A., Evans,
N.~J., II, \& Martel, H.,\ 2010b, in preparation
\bibitem[Vazquez-Semadeni(1994)]{1994ApJ...423..681V} Vazquez-Semadeni, E.\ 
1994, \apj, 423, 681 
\bibitem[V{\'a}zquez-Semadeni 
\& Garc{\'{\i}}a(2001)]{2001ApJ...557..727V} V{\'a}zquez-Semadeni, E., 
\& Garc{\'{\i}}a, N.\ 2001, \apj, 557, 727 
\bibitem[Wong et al.(2008)]{wong08} Wong, T., et al.\ 2008, 
\mnras, 386, 1069 

 

\end{thebibliography}
\end{document}